\documentclass{article}

\font\scap=cmcsc10 \hfuzz=5cm
\font\scap=cmcsc10

\usepackage{graphicx}

\newcount\eqnumber
\def\neweq{{\rm{(\the\eqnumber)}}\global\advance\eqnumber by 1}
\def\eqdef#1{\eqno\xdef#1{\the\eqnumber}\neweq}
\def\newaeq{{\rm{(\the\eqnumber {\it a})}}\global\advance\eqnumber by 1}
\def\eqdaf#1{\eqno\xdef#1{\the\eqnumber}\newaeq}
\def\eqdisp#1{\xdef#1{\the\eqnumber}\neweq}
\def\eqdasp#1{\xdef#1{\the\eqnumber}\newaeq}

\def\newaleq{{\rm{\the\eqnumber}}\global\advance\eqnumber by 1}
\def\eqdal#1{\xdef#1{\the\eqnumber}\newaleq}

\newcount\refnumber
\def\newref{{\the\refnumber}\global\advance\refnumber by 1}
\def\refdef#1{{\xdef#1{\the\refnumber}}\newref}

\begin{document}

\centerline{\bf Discrete Painlev\'e equations from singularity patterns: the asymmetric trihomographic case}

\bigskip
\bigskip{\scap B. Grammaticos} and {\scap A. Ramani}
\quad{\sl IMNC,  CNRS, Universit\'e Paris-Diderot, Universit\'e Paris-Sud, Universit\'e Paris-Saclay, 91405 Orsay, France}

\medskip{\scap R. Willox}\quad
{\sl Graduate School of Mathematical Sciences, the University of Tokyo, 3-8-1 Komaba, Meguro-ku, 153-8914 Tokyo, Japan }

\medskip{\scap J. Satsuma}
\quad{\sl  Department of Mathematical Engineering, Musashino University, 3-3-3 Ariake, Koto-ku, 135-8181 Tokyo, Japan}

\bigskip
{\scap Abstract}
\smallskip

We derive the discrete Painlev\'e equations associated to the affine Weyl group E$_8^{(1)}$ that can be represented by an  (in the QRT sense) ``asymmetric'' trihomographic system. The method used in this paper is based on singularity confinement. We start by obtaining all possible singularity patterns for a general asymmetric trihomographic system and discard those patterns which cannot lead to confined singularities. Working with the remaining ones we implement the confinement conditions and derive the corresponding discrete Painlev\'e equations, which involve two variables. By eliminating either of these variables we obtain a ``symmetric'' equation. Examining all these equations of a single variable, we find that they coincide exactly with those derived in previous works of ours, thereby establishing the completeness of our results. 

\bigskip

\bigskip
1. {\scap Introduction}
\medskip

There exists a plethora of methods for the derivation of discrete Painlev\'e equations [\refdef\dps]. These belong to five major classes.

a) Equations obtained from some other Painlev\'e equation. The general approach is to use the Schlesinger transformations for continuous Painlev\'e equations in order to construct discrete ones. The best known example is that of the contiguity relations of continuous Painlev\'e equations, which are non-autonomous systems that are integrable by construction and which turn out to be discrete Painlev\'e equations. A classic case is that of the contiguity of the Painlev\'e II equation obtained by Jimbo and Miwa [\refdef\jimbo]
$${z_{n+1}\over x_{n+1}+x_n}+{z_{n}\over x_{n}+x_{n-1}}=-x_n^2+1,\eqdef\zena$$
where $z_n=\alpha n+\beta$, which was (later) recognised [\refdef\fokas] as an alternative discrete form of Painlev\'e I. 

Some contiguity relations for discrete Painlev\'e equations do not yield new results due to the property of self-duality [\refdef\selfdu] which holds for most discrete Painlev\'e equations, except for those associated with the affine Weyl groups [\refdef\sakai]  A$_2^{(1)}$+A$_1^{(1)}$ and A$_1^{(1)}$+A$_1^{(1)}$. However, for systems with two or more parameters it is always possible to obtain contiguity relations by considering different directions of evolution.

Another possibility which exists is that of the Miura relations [\refdef\miura] between discrete Painlev\'e equations. Not only does the Miura relation allow one to obtain a new equation starting from some discrete Painlev\'e equation, but the Miura system itself also constitutes an equation in its own right. 

b) Equations obtained by the reduction of some higher-dimensional system.  This is in perfect parallel to the continuum situation where all (continuous) Painlev\'e equations can be obtained as one-dimensional reduction of some two-dimensional integrable evolution equation. A well-known example of such a contruction is the derivation by Nijhoff and Papageorgiou [\refdef\fravas]. Starting from an integrable lattice of KdV type and performing a discrete similarity reduction they obtained the mapping
$$x_{n+1}+x_{n-1}={z_nx_n+a\over 1-x_n^2},\eqdef\zdyo$$
which, in perfect parallel to the continuum case, is a discrete analogue of Painlev\'e II. 

c) Equations obtained starting from some inverse problem. Examples of the latter are recursions involving orthogonal polynomials, the discrete AKNS method [\refdef\akns], the methods of discrete dressing [\refdef\disdres], of non-isospectral deformations and so on. In particular, the method of orthogonal polynomials is linked to the genesis of discrete Painlev\'e equations since, already in 1939, Shohat [\refdef\shohat] used it in order to derive the integrable non-autonomous recursion relation
$$x_{n+1}+x_n+x_{n-1}={z_n\over x_n}+1,\eqdef\ztri$$
$z_n=\alpha n+\beta+\gamma(-1)^n$. 
This recursion relation resurfaced (much later) in the work of Brezin and Kazakov [\refdef\brezov] where it was identified (when $\gamma=0$) as the discrete analogue of Painlev\'e I.

d)  Equations obtained from the geometry of some affine Weyl group. This method is based on the Sakai classification [\sakai]. As  Sakai has shown, the discrete Painlev\'e equations can be associated to affine Weyl groups, the latter  forming a degeneration pattern starting from the group E$_8^{(1)}$. 

\vskip.3cm
\centerline{{\includegraphics[width=15cm,keepaspectratio]{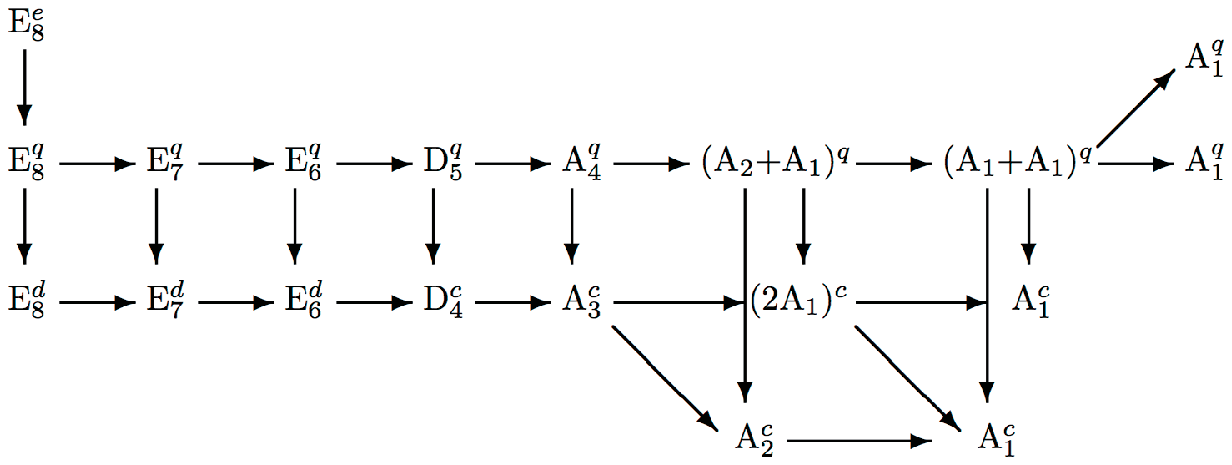}}}

This pattern shows all possible degenerations, starting from the elliptic discrete Painlev\'e equation all the way to the zero-parameter ones, explicitly including degenerations to multiplicative- and additive-type equations. The upper indices $e,q,d,c$ appearing in the names of the groups refer to the type of equations encountered in each of them, namely elliptic, multiplicative, difference and equations which are contiguity relations of continuous Painlev\'e equations [\refdef\kanoya].

One establishes the basic Miura relation and uses it in order to construct discrete Painlev\'e equations associated to a chosen group in the cascade. We find, for instance, for the additive E$_8^{(1)}$-associated discrete Painlev\'e equation [\refdef\eight] the form
$${(y_n-x_{n+1}+(\zeta_n+z_{n+1})^2)(y_n-x_{n}+(\zeta_n+z_n)^2)+4y_n(\zeta_n+z_{n+1})(\zeta_n+z_n)\over 
(\zeta_n+z_n)(y_n-x_{n+1}+(\zeta_n+z_{n+1})^2)+(\zeta_n+z_{n+1})(y_n-x_{n}+(\zeta_n+z_n)^2)}
=2{y_n^4+S_2y_n^3+S_4y_n^2+S_6y_n+S_8\over S_1y_n^3+S_3y_n^2+S_5y_n+S_7},\eqdaf\ztes$$
$${(x_n-y_n+(z_n+\zeta_{n})^2)(x_n-y_{n-1}+(z_n+\zeta_{n-1})^2)+4x_n(z_n+\zeta_{n})(z_n+\zeta_{n-1})\over 
(z_n+\zeta_{n-1})(x_n-y_n+(z_n+\zeta_{n})^2)+(z_n+\zeta_{n})(x_n-y_{n-1}+(z_n+\zeta_{n-1})^2)}
=2{x_n^4+\tilde S_2x_n^3+\tilde S_4x_n^2+\tilde S_6x_n+\tilde S_8\over \tilde S_1x_n^3+\tilde S_3x_n^2+\tilde S_5x_n+\tilde S_7},\eqno(\ztes b)$$
where $z_n=t_n$ and $\zeta_n=t_n+\alpha/2$ and $S_k,\tilde S_k$ are the elementary symmetric functions of the quantities $z_n+\kappa_n^i$ and $\zeta_n-\kappa_n^i$, where $\kappa^i$ are eight parameters. 

e) Equations obtained through the deautonomisation procedure. The latter is a method we introduced (albeit not under that moniker) in our very first paper [\dps] on discrete Painlev\'e equations and which has been massively used since for the derivation of these systems. The deautonomisation method consists in extending an autonomous mapping to one where the various, previously constant, coefficients are (appropriately chosen) functions of the independent variable. Based on an analogy to the differential case the choice of the initial autonomous mapping was almost invariably one belonging to the QRT [\refdef\qrt] family. (The analogy we refer to here, has to do with the fact that the Painlev\'e equations are nonautonomous extensions of equations which are solved by elliptic functions which is also true for the QRT mappings. Thus we expect that the nonautonomous extension of the latter will lead to discrete Painlev\'e equations).

The practical application of the deautonomisation is through the use of a discrete integrability criterion, typically the singularity confinement property [\refdef\sincon], i.e. we use the latter is order to select the appropriate non-autonomous extensions of the previously constant coefficients. The justification of the deautonomisation procedure based on an algebro-geometric approach was presented in [\refdef\procroy], a study which led to the proposal of a stronger version of the singularity confinement criterion [\refdef\full] under the name of `full deautonomisation'. 

This paper will be devoted to the study of discrete Painlev\'e equations associated to the affine Weyl group E$_8^{(1)}$. The approach we shall use here is based on a canonical form we introduced a while back and which we dubbed `trihomographic' [\refdef\trihom]. In some previous papers of ours, we investigated the existence of discrete Painlev\'e equations that can be written as (in the QRT sense) asymmetric,  trihomographic, equations. In particular, in [\refdef\first] we started from two already known results of ours which had singularities confining after (8,6,1,1) and (12,2,1,1), for a total of 16 steps. We argued there that if such singularity patterns existed there was no reason why confined patterns involving (10,4,1,1), (6,4,4,2) and (4,4,4,4) steps would not exist. This was borne out by our analysis and the corresponding discrete Painlev\'e equations  were derived. However the approach we adopted there was mainly heuristic. While it did lead to new results, it did not preclude the existence of more equations and the development of a systematic approach was therefore in order. This is the topic of the present paper. As before, we shall work with asymmetric trihomographic systems but here we shall examine {\sl all possible} singularity patterns and, as we shall see,  many more discrete Painlev\'e equations do exist, associated to new, not previously considered singularity patterns. 

\bigskip
2. {\scap The trihomographic and ancillary representations}
\medskip

The trihomographic representation was introduced based on the form of the elementary Miura transformation obtained from the geometrical description [\eight] of equations associated to the group E$_8^{(1)}$. The latter relates three variables $X,Y,Z$ and and is given by the expression
$${X-A\over X-B}{Y-C\over Y-D}{Z-E\over Z-F}=G.\eqdef\zpen$$
Inspired by the form of this expression, we introduced a simple symmetric form tailored to additive equations
$${x_{n+1}-(z_{n}+k_n)^2\over x_{n+1}-(z_{n}-k_n)^2}\ {x_{n-1}-(z_{n+1}+k_n)^2\over x_{n-1}-(z_{n+1}-k_n)^2}\ {x_{n}-(z_{n+1}+z_n-k_n)^2\over x_{n}-(z_{n+1}+z_n+k_n)^2}=1,\eqdef\zhex$$
where $z_n, k_n$ are, as yet unspecified, functions of the independent variable $n$. (As shown in [\refdef\second], once the additive trihomographic form of an E$_8^{(1)}$-associated equation is obtained, the extension to the multiplicative and elliptic cases is straightforward). Extending (\zhex) to an asymmetric form (in the QRT sense)  leads to 
$${x_{n+1}-(z_{n}+k_n)^2\over x_{n+1}-(z_{n}-k_n)^2}\ {x_{n}-(\zeta_n+k_n)^2\over x_{n}-(\zeta_n-k_n)^2}\ {y_{n}-(\zeta_n+z_n-k_n)^2\over y_{n}-(\zeta_n+z_n+k_n)^2}=1\eqdaf\zhep$$
$${y_{n}-(\zeta_{n-1}+\kappa_n)^2\over y_{n}-(\zeta_{n-1}-\kappa_n)^2}\ {y_{n-1}-(z_{n}+\kappa_n)^2\over y_{n-1}-(z_{n}-\kappa_n)^2}\ {x_{n}-(\zeta_{n-1}+z_n-\kappa_n)^2\over x_{n}-(\zeta_{n-1}+z_n+\kappa_n)^2}=1,\eqno(\zhep b)$$
where $\zeta_n,\kappa_n$ are as yet unspecified, just like $z_n, k_n$.
This trihomographic form is equivalent to the generic additive form, for a specific right-hand side. We have indeed
$${(y_n-x_{n+1}+\zeta_n^2)(y_n-x_{n}+z_n^2)+4y_n\zeta_nz_n\over z_n(y_n-x_{n+1}+\zeta_n^2)+\zeta_n(y_n-x_{n}+z_n^2)}
={y_n-k_n^2\over z_n+\zeta_n}+z_n+\zeta_n,\eqdaf\zoct$$
$${(x_n-y_n+z_n^2)(x_n-y_{n-1}+\zeta_{n-1}^2)+4x_nz_n\zeta_{n-1}\over 
\zeta_{n-1}(x_n-y_n+z_n^2)+z_n(x_n-y_{n-1}+\zeta_{n-1}^2)}
={x_n-\kappa_n^2\over z_n+\zeta_{n-1}}+z_n+\zeta_{n-1}.\eqno(\zoct b)$$
Note that the notation used in (\zoct) is slightly different from that in (\ztes): the quantities $z_n+\zeta_n$ and $z_{n+1}+\zeta_n$ in (\ztes) are represented in (\zoct) by $z_n$ and $\zeta_n$,  leading to more compact expressions. We must point out here that the equivalence between the trihomographic and the canonical forms is a formal one and holds independently of the precise expressions of the parameters $z,\zeta, k,\kappa$. 

The application of singularity analysis based on trihomographic forms led to the results referred to at the end of the introduction. However, as we pointed out there a more systematic approach is in order. For this purpose, we shall work with the representation proposed in [\refdef\ancil] and which is based on the introduction of an ancillary variable. For the additive equations we are focusing on in this paper, the latter is obtained from
$$x_n=\xi_n^2\quad {\rm and,\ in\ the\ case\ of\ asymmetric\ systems,} \quad y_n=\eta_n^2.\eqdef\zenn$$
This, very convenient, ancillary variable for the additive equation associated to the affine Weyl group $E_8^{(1)}$ was also independently obtained by Kajiwara, Noumi and Yamada in [\kanoya]. (Ancillary variables can also be introduced for multiplicative and elliptic equations, as we showed in [\refdef\multell]).
Using these ancillary variables we can rewrite the general additive E$_8^{(1)}$- associated equation as
$${x_{n+1}-(\eta_n-\zeta_n)^2\over x_{n+1}-(\eta_n+\zeta_n)^2}\,{x_{n}-(\eta_n-z_n)^2\over x_{n}-(\eta_n+z_n)^2}={\prod_{i=1}^8(\eta_n-C_i)\over\prod_{i=1}^8(\eta_n+C_i)}\eqdaf\zdek$$
$${y_{n}-(\xi_n-z_n)^2\over y_{n}-(\xi_n+z_n)^2}\,{y_{n-1}-(\xi_n-\zeta_{n-1})^2\over y_{n-1}-(\xi_n+\zeta_{n-1})^2}={\prod_{i=1}^8(\xi_n-A_i)\over\prod_{i=1}^8(\xi_n+A_i)},\eqno(\zdek b)$$
where $A_i$ and $C_i$ are parameters related to the $\kappa_i$ of (\ztes).
We remark that both the left and right hand sides of (\zdek) are expressed in a factorised form thanks to the introduction of the ancillary variables $\xi,\eta$. This makes the application of singularity analysis particularly convenient. Let us show how this works in the case of the generic additive E$_8^{(1)}$ equation. The generic character of the equation means that all singularities are confined in just one step. Entering a singularity by, say, $\xi_n=A_i(n)$ we find $y_n=(A_i(n)-z_n)^2$ or, without loss of generality, $\eta_n=A_i(n)-z_n$, and the singularity is confined provided $A_i(n)+C_i(n)=z_n$. (A permutation of the exit points, i.e. the $C_i$'s, could have been considered, but, as we explained in [\ancil] this would have introduced fictitious periodic dependencies, removable by an adequate gauge tranformations).  Similarly, entering the singularity by $\eta_n=C_i(n)$ we find the confinement constraint $A_i(n+1)+C_i(n)=\zeta_n$.  Another possibility for a singularity may arise whenever $x$ or $y$ take the value $\infty$. Requiring that the two members of the equations balance, so that an infinite value of $x$ or $y$ is, in fact,  not a singularity, we find the constraints
$$\sum_{i=1}^8 A_i=2(z_n+\zeta_{n-1})\quad {\rm and}\quad \sum_{i=1}^8 C_i=2(z_n+\zeta_{n}).\eqdef\dena$$
 Using the confinement constraints $A_i(n)+C_i(n)=z_n$, $A_i(n+1)+C_i(n)=\zeta_n$ together with (\dena) we find that $z$ obeys the equation $z_{n+1}-2z_n+z_{n-1}=0$, i.e. $z_n=\alpha_n+\beta$ and $\zeta$ is related to $z$ by $\zeta_n=(z_{n+1}+z_n)/2$.
 
The trihomographic form can, obviously, be recast in a form involving the ancillary variables. Starting from (\zhep) we find indeed the system
$${x_{n+1}-(\eta_n-\zeta_n)^2\over x_{n+1}-(\xi_n+\zeta_n)^2}\,{x_{n}-(\eta_n-z_n)^2\over x_{n}-(\eta_n+z_n)^2}={(\eta_n-C_n)(\eta_n-D_n)\over (\eta_n+C_n)(\eta_n+D_n)}\eqdaf\ddyo$$
$${y_{n}-(\xi_n-z_n)^2\over y_{n}-(\xi_n+z_n)^2}\,{y_{n-1}-(\xi_n-\zeta_{n-1})^2\over y_{n-1}-(\xi_n+\zeta_{n-1})^2}={(\xi_n-A_n)(\xi_n-B_n)\over(\xi_n+A_n)(\xi_n+B_n)},\eqno(\ddyo b)$$
where the $A,B,C,D$ are related to the paramaters $k$ and $\kappa$ of (\zhep) by $A_n=z_n+\zeta_{n-1}+\kappa_n$,  $B_n=z_n+\zeta_{n-1}-\kappa_n$, $C_n=z_n+\zeta_{n}+k_n$, and $D_n=z_n+\zeta_{n}-k_n$. Note that the choice of signs for $k_n, \kappa_n$ is immaterial: a change of these signs would correspond to a global inversion of either of the left-hand sides of (\zhep). Clearly the $A,B,C,D$ obey the constraint for the non-existence of the singularity at infinity:
$$A_n+B_n=2(z_n+\zeta_{n-1})\quad {\rm and}\quad C_n+D_n=2(z_n+\zeta_{n}).\eqdef\dtri$$
Suppose now that we enter a singularity through $\xi_n=A_n$. We find, as in the generic case, that $\eta_n=A_n-z_n$ and (unless the singularity is confined at this step) $\xi_{n+1}=A_n-z_n-\zeta_n$ and $\eta_{n+1}=A_n-z_n-\zeta_n-z_{n+1}$. Thus, pursuing the evolution of the singularity we find that each step subtracts a  $\zeta$ from $\xi$ and a $z$ from $\eta$ respectively, with the appropriate indices. Similarly, when we enter a singularity through $\eta_n=C_n$ we find $\xi_{n+1}=C_n-\zeta_n$ and $\eta_{n+1}=C_n-\zeta_n-z_{n+1}$,  $\xi_{n+2}=C_n-\zeta_n-z_{n+1}-\zeta_{n+1}$. For the singularity to confine, the value of $\xi$ or $\eta$ after a number of steps ,must be equal to the value of one of the $A,B,C,D$ leading to a relation of the form $E_n+F_{n+k}$ equal to a sum of $z,\zeta$ at the appropriate indices (where $E,F$ stand for any of the $A,B,C,D$). 

The singularity analysis we are going to present in the next sections is based on a systematic exploration of all possible singularity patterns. Namely, we shall require that each of the entry points, i.e. $\xi_n=A_n\, {\rm or}\ B_n$ and $\eta_n=C_n\, {\rm or}\ D_n$, exits through some of the $A,B,C,D$ in a given number of steps. Thus a singularity pattern is represented by these four numbers of necessary steps, $(M,N,P,Q)$, where the sum of the four is always equal to 16. As we have seen in the introduction there exist systems, already identified in [\first], with patterns (12,2,1,1), (10,4,1,1), (8,6,1,1), (6,4,4,2), and (4,4,4,4). Here we shall consider all the patterns comprised of the quartet of positive integers $(M,N,P,Q)$ with $M+N+P+Q=16$ and identify those that indeed correspond to a confined singularity when the system is deautonomised. 

\bigskip
3. {\scap Singularity analysis of the all-even or all-odd steps case}
\medskip

The first set of possibilities we shall examine is when all $M,N,P,Q$ are even. Practically this means that if one enters a singularity through $A$ or $B$, one must exit it through either $A$ or $B$ in $M,N$ steps, and similarly for $C$ and $D$. The only a priori possible patterns are (10,2,2,2), (8,4,2,2), (6,6,2,2), (6,4,4,2), and (4,4,4,4).  However not all of them can exist. In order to see this it suffices to consider the autonomous case where $A,B,C,D, z,\zeta$ are constant. The constraints (\dtri) become now $A+B=2(z+\zeta)$ and $C+D=2(z+\zeta)$ and, assuming that $z+\zeta$ is not zero we can normalise it to 2. Requiring that $A$ exits through $A$ or $B$ in $M$ steps and similarly for $B$ exiting through $A$ or $B$ in $N$ steps we find a relation of the form $A+E=M$ and $B+F=N$, where $E,F$ are equal to either $A,B$ or $B,A$. Adding the two relations we find that $2(A+B)=M+N$. But from (\dtri) we have $A+B=4$ and thus $M+N=8$, and similarly $P+Q=8$. Thus clearly the patterns (10,2,2,2), (8,4,2,2) are incompatible with the confinement constraints and we are left with (6,6,2,2), (6,4,4,2), and (4,4,4,4), but the order of the lengths, in the case of the first two, is still free.

Three classes of singularity patterns are possible. The first corresponds to the pattern $\{A\to A, B\to B, C\to C,D\to D\}$, where by $\{ E\to F\}$ we mean that we are entering the singularity through $E$ and exiting it through $F$. The second class is $\{A\to A, B\to B, C\to D, D\to C\}$ (without loss of generality) and the third one $\{A\to B, B\to A, C\to D, D\to C\}$. The pattern lengths $\{6,2,6,2\}$ can only exist for the first class. In fact it corresponds to a symmetric equation with a singularity pattern (6,2), an equation already obtained in [\second] as case I, which is here written in asymmetric form by artificially doubling the number of variables. The pattern $\{6,2,4,4\}$ exists for both the first and the second classes. Again these results are not new: they correspond to cases XI and X of [\second] respectively. Finally the pattern $\{4,4,4,4\}$ is possible for all three classes, corresponding to cases obtained in [\second], V, XII and II respectively. Note that the first and third cases are in fact symmetric, here artificially cast into asymmetric form.

An interesting remark concerns the systems $\{6,2,6,2\}$ and $\{4,4,4,4\}$. The fact that these can be written in asymmetric form by artificially doubling the number of variables means that considering the equation obtained by elimination of one of the variables is tantamount to skipping one step out of two in the evolution, i.e. to  considering the double-step evolution. Indeed, by eliminating either of the two variables in the two systems we obtain two equations first derived in [\ancil], namely equations  4.2.1 or 5.2.1. And precisely these two equations have been obtained as double-step evolutions in  [\refdef\eaone] and [\refdef\restor] respectively. 

Thus the all-even steps case does not lead to new discrete Painlev\'e equations. 

The second set we analyse is when all $M,N,P,Q$ are odd (and we shall take $M=2m+1$ and analogously for $N, P$ and $Q$). This means that if one enters the singularity through $A$ or $B$ one must exit it through either $C$ or $D$ in $M,N$ steps, and similarly for $C$ and $D$. Here the patterns that are, a priori, possible are (13,1,1,1), (11,3,1,1), (9,5,1,1), (9,3,3,1), (7,7,1,1), (7,5,3,1), (7,3,3,3), (5,5,5,1), and (5,5,3,3). Having an odd number of steps means that $A$ exits through $C$ (without loss of generality) and $B$ though $D$. Similarly $C,D$ exit through $A,B$ or $B,A$, defining thus two different classes of singularity patterns, namely $\{A\to C, B\to D, C\to A,D\to B\}$ and $\{A\to C, B\to D, C\to B,D\to A\}$.

Considering the autonomous case we have for the first class the constraints: $A+C=(m+1)z+m\zeta$, $B+D=(n+1)z+n\zeta$, $C+A=pz+(p+1)\zeta$, $D+B=qz+(q+1)\zeta$. Adding the first and last relations, and using the constraint coming from (\dtri) we arrive at the constraint $M+Q=8$. Thus the first class can exist only in the case of the patterns (7,7,1,1), (7,5,3,1), and (5,5,3,3). Unfortunately no constraint exists for the second class patterns and thus all possible lengths must be examined. 

In order to show how the singularity analysis is implemented let us analyse in detail the case of the pattern (7,7,1,1). First we remark that the first class can be realised in two different ways, corresponding to patterns $\{7,7,1,1\}$ and $\{7,1,7,1\}$, where the display of the quartet of steps within braces indicates the order in which the lengths fo the singularities appear.
The second class can be realised through patterns $\{7,7,1,1\}$ and $\{7,1,7,1\}$, while the third possible pattern $\{7,1,1,7\}$ is identical to $\{7,1,7,1\}$ under exchange of $A,C$ and $B,D$ and inversion of the direction of evolution. 
\medskip
For pattern $\{A\to C, B\to D, C\to A,D\to B\}$ with $\{7,7,1,1\}$ we find:
$$A_n+C_{n+3}=z_n+\zeta_n+z_{n+1}+\zeta_{n+1}+z_{n+2}+\zeta_{n+2}+z_{n+3}\eqdaf\dtes$$
$$B_n+D_{n+3}=z_n+\zeta_n+z_{n+1}+\zeta_{n+1}+z_{n+2}+\zeta_{n+2}+z_{n+3}\eqno(\dtes b)$$
$$ C_n+A_{n+1}=\zeta_n\eqno(\dtes c)$$
$$ D_n+B_{n+1}=\zeta_n.\eqno(\dtes d)$$
The integration of these constraints leads to:
$$z_n=-2(\alpha n+\beta)+\phi_2(n)+\phi_3(n)$$
$$\zeta_n=4(\alpha n+\beta)+\alpha+\phi_3(n-1)$$
$$A_n=2(\alpha n+\beta)-2\alpha+\phi_2(n)-\phi_3(n-1)+\phi_4(n)+\gamma$$
$$B_n=2(\alpha n+\beta)-2\alpha+\phi_2(n)-\phi_3(n-1)-\phi_4(n)-\gamma$$
$$C_n=2(\alpha n+\beta)+2\alpha+\phi_2(n)-\phi_3(n+1)-\phi_n(n+1)-\gamma$$
$$D_n=2(\alpha n+\beta)+2\alpha+\phi_2(n)-\phi_3(n+1)+\phi_4(n+1)+\gamma,$$
where $\phi_m$ is a periodic functions with period $m$, given  by
$$ \phi_m(n)=\sum_{l=1}^{m-1} \delta_l^{(m)} \exp\left({2i\pi ln\over m}\right),\eqdef\dpen$$
Note that the summation starts at 1 instead of 0 and thus $\phi_m$ introduces $m-1$ parameters.

Eliminating either $x$ or $y$ from (\ddyo) in this case leads to a trihomographic equation identified in [\second] as case II.

The pattern $\{A\to C, B\to D, C\to A,D\to B\}$ with $\{7,1,7,1\}$ is just an artificial doubling of the number of variables, in the case III obtained in [\second].  Eliminating either of the two variable in this case leads to equation 4.3.4 of [\ancil], which, as shown in [\refdef\sustem] corresponds to a double-step evolution  starting from case III.

\smallskip
For pattern $\{A\to C, B\to D, C\to B,D\to A\}$ with $\{7,7,1,1\}$ we find,
$$A_n+C_{n+3}=z_n+\zeta_n+z_{n+1}+\zeta_{n+1}+z_{n+2}+\zeta_{n+2}+z_{n+3}\eqdaf\dhex$$
$$ B_n+D_{n+3}=z_n+\zeta_n+z_{n+1}+\zeta_{n+1}+z_{n+2}+\zeta_{n+2}+z_{n+3}\eqno(\dhex b)$$
$$C_n+B_{n+1}=\zeta_n\eqno(\dhex c)$$
$$ D_n+A_{n+1}=\zeta_n.\eqno(\dhex d)$$
The integration of these constraints leads to:
$$z_n=-2(\alpha n+\beta)+\phi_2(n)+\phi_3(n)$$
$$\zeta_n=4(\alpha n+\beta)+2\alpha+\phi_3(n-1)$$
$$A_n=2(\alpha n+\beta)-2\alpha+\phi_2(n)-\phi_3(n-1)+\chi_8(n)$$
$$B_n=2(\alpha n+\beta)-2\alpha+\phi_2(n)-\phi_3(n-1)-\chi_8(n)$$
$$C_n=2(\alpha n+\beta)+2\alpha+\phi_2(n)-\phi_3(n+1)+\chi_8(n+1)$$
$$D_n=2(\alpha n+\beta)+2\alpha+\phi_2(n)-\phi_3(n+1)-\chi_8(n+1),$$
where $\chi_{2m}$ is a periodic function with period $2m$. It is given  
$$ \chi_{2m}(n)=\sum_{\ell=1}^{m} \eta_{\ell}^{(m)} \exp\left({i\pi(2\ell-1)n\over m}\right),\eqdef\dhep$$
introducing $m$ free parameters.

Eliminating either $x$ or $y$ from (\ddyo) in this case leads to a trihomographic equation identified in [\second] as case V.

\smallskip
Finally we have the pattern $\{A\to C, B\to D, C\to B,D\to A\}$ with $\{7,1,7,1\}$.

The confinement constraints are
$$A_n+C_{n+3}=z_n+\zeta_n+z_{n+1}+\zeta_{n+1}+z_{n+2}+\zeta_{n+2}+z_{n+3}\eqdaf\doct$$
$$B_n+D_{n}=z_n\eqno(\doct b)$$
$$ C_n+B_{n+4}=\zeta_n+z_{n+1}+\zeta_{n+1}+z_{n+2}+\zeta_{n+2}+z_{n+3}+\zeta_{n+3}\eqno(\doct c)$$
$$D_n+A_{n+1}=\zeta_n,\eqno(\doct d)$$
the integration of which leads to:
$$z_n=\alpha n+\beta+\phi_2(n)+\phi_3(n)+\chi_8(n)$$
$$\zeta_n=\alpha n+\beta-4\alpha+\phi_2(n)+\phi_3(n-1)-\chi_8(n)$$
$$A_n=2(\alpha n+\beta-4\alpha)+2\phi_2(n)-\phi_3(n-1)+\chi_8(n)+\chi_8(n+2)-\chi_8(n-1)-\chi_8(n+1)$$
$$B_n=2(\alpha n+\beta-\alpha)-2\phi_2(n)-\phi_3(n-1)+\chi_8(n)-\chi_8(n+2)-\chi_8(n-1)+\chi_8(n+1)$$
$$C_n=5(\alpha n+\beta-2\alpha)+\phi_2(n)-\phi_3(n+1)-\chi_8(n+2)-\chi_8(n-1)+\chi_8(n+1)$$
$$D_n=-\alpha n-\beta+2\alpha+3\phi_2(n)-\phi_3(n+1)+\chi_8(n+2)+\chi_8(n-1)-\chi_8(n+1).$$
Eliminating $y$ from (\ddyo) in this case leads to a trihomographic equation for $x$ identified in [\second] as case V. On the other hand, eliminating $x$ leads for $y$ to an equation derived in [\ancil], case 4.3.3.

In what follows we shall not explicitly give the confinement constraints: writing them once the pattern and the singularity steps are given is elementary. Thus we shall limit ourselves to the results for the various quantities entering the equation.
\medskip
For the pattern $\{A\to C, B\to D, C\to B,D\to A\}$ with $\{13,1,1,1\}$,
we find

$$z_n=-2(\alpha n+\beta)+\alpha +\phi_3(n)+\phi_5(n)$$
$$\zeta_n=4(\alpha n+\beta)+\phi_3(n-1)-\phi_5(n)-\phi_5(n+1)$$
$$A_n=5(\alpha n+\beta)-6\alpha+\phi_2(n)-\phi_3(n-1)+\phi_5(n+2)-\phi_5(n-1)$$
$$B_n=-(\alpha n+\beta)-\phi_2(n)-\phi_3(n-1)-\phi_5(n+2)-\phi_5(n-1)$$
$$C_n=5(\alpha n+\beta)+\alpha-\phi_2(n)-\phi_3(n+1)-\phi_5(n-2)-\phi_5(n+1)$$
$$D_n=-(\alpha n+\beta)+\alpha+\phi_2(n)-\phi_3(n+1)+\phi_5(n-2)-\phi_5(n+1).$$

Eliminating either $x$ or $y$ from (\ddyo) in this case leads to a trihomographic equation identified in [\second] as case I.
\medskip
For the pattern $\{A\to C, B\to D, C\to B,D\to A\}$ with (11,3,1,1), two cases should be distinguished here, corresponding to the step sequences $\{11,3,1,1\}$ and $\{11,1,3,1\}$, the sequence $\{11,1,1,3\}$ being equivalent to the latter of the two after an exchange of $A,C$ and $B,D$ and a reversal of the evolution direction. 

In the first case we find the solution
$$z_n=-2(\alpha n+\beta)+\phi_2(n)+\phi_5(n)$$
$$\zeta_n=4(\alpha n+\beta)+2\alpha-\phi_5(n)-\phi_5(n+1)$$
$$A_n=2(\alpha n+\beta)-3\alpha+\phi_2(n)+\phi_5(n)+\phi_5(n-2)+\phi_5(n+1)+\chi_4(n)$$
$$B_n=-\alpha+\phi_2(n)+\phi_5(n+2)-\phi_5(n-1)-\chi_4(n)$$
$$C_n=2(\alpha n+\beta)+3\alpha+\phi_2(n)-\phi_5(n-2)-\phi_5(n+1)+\chi_4(n+1)$$
$$D_n=\alpha+\phi_2(n)+\phi_5(n-2)-\phi_5(n+1)-\chi_4(n+1).$$

Eliminating either $x$ or $y$ from (\ddyo) in this case leads to the equation identified in [\second] as case I.

In the second case we obtain
$$z_n=-(\alpha n+\beta)+\phi_4(n)+\phi_5(n)$$
$$\zeta_n=3(\alpha n+\beta)+4\alpha+\phi_4(n+2)+\phi_5(n-1)+\phi_5(n+2)$$
$$A_n=4(\alpha n+\beta)+\alpha+\phi_4(n)-\phi_4(n-1)-2\phi_5(n-1)-\phi_5(n+2)-\phi_5(n+1)$$
$$B_n=\alpha+\phi_4(n+1)-\phi_4(n+2)-\phi_5(n+2)+\phi_5(n+1)$$
$$C_n=5(\alpha n+\beta)+9\alpha-\phi_4(n-1)+\phi_5(n+2)+\phi_5(n-1)$$
$$D_n=-(\alpha n+\beta)-\alpha-\phi_4(n-1)-2\phi_4(n+1)+\phi_5(n+2)-\phi_5(n+1)+\phi_5(n).$$
Eliminating $y$ from (\ddyo) in this case leads again to case I, 
while eliminating $x$ leads for $y$ to the equation identified as case 4.5.2 in [\ancil].
\medskip
For the pattern $\{A\to C, B\to D, C\to B,D\to A\}$ with (9,5,1,1), again two cases must be distinguished, corresponding to the step sequences $\{9,5,1,1\}$ and $\{9,1,5,1\}$ (and $\{9,1,1,5\}$ is equivalent to the latter just as in the previous cases).

In the first case we find
$$z_n=-2(\alpha n+\beta)+\alpha+\phi_7(n)$$
$$\zeta_n=4(\alpha n+\beta)-\phi_7(n)-\phi_7(n+1)$$
$$A_n=3(\alpha n+\beta)-4\alpha+\phi_2(n)-\phi_7(n+3)+\phi_7(n+1)+\phi_7(n)+\phi_7(n-2)$$
$$B_n=\alpha n+\beta-2\alpha-\phi_2(n)+2\phi_7(n+3)+\phi_7(n+2)-\phi_7(n-1)+\phi_7(n-3)$$
$$C_n=3(\alpha n+\beta)+\alpha-\phi_2(n)+\phi_7(n+2)+\phi_7(n)+\phi_7(n-1)-\phi_7(n-3)$$
$$D_n=\alpha n+\beta+\alpha+\phi_2(n)+\phi_7(n+3)-\phi_7(n+1)+\phi_7(n-2)+2\phi_7(n-3).$$

Eliminating either $x$ or $y$ from (\ddyo) in this case leads to a trihomographic equation identified in [\second] as case IV.

The second case leads to
$$z_n=2\alpha+\phi_7(n)-\phi_7(n-1)$$
$$\zeta_n=2(\alpha n+\beta)+\alpha +\phi_7(n+2)-\phi_7(n+1)+\phi_7(n-1)$$
$$A_n=3(\alpha n+\beta)-\alpha+\phi_2(n)-\phi_7(n+3)-2\phi_7(n-1)+\phi_7(n-2)$$
$$B_n=\alpha n+\beta+3\alpha-\phi_2(n)+\phi_7(n+3)+2\phi_7(n+1)+\phi_7(n-2)$$
$$C_n=5(\alpha n+\beta)+7\alpha-\phi_2(n)+\phi_7(n+2)-\phi_7(n+1)-\phi_7(n-3)$$
$$D_n=-(\alpha n+\beta)-\alpha+\phi_2(n)+\phi_7(n+2)-\phi_7(n+1)+2\phi_7(n)+\phi_7(n-3).$$

Eliminating either $x$ or $y$ from (\ddyo) in this case leads to an equation obtained in [\ancil], case 4.4.3.
\medskip
For pattern $\{A\to C, B\to D, C\to B,D\to A\}$ with (9,3,3,1) there are also two distinct cases, corresponding to $\{9,3,3,1\}$ and $\{9,1,3,3\}$ (and obviously $\{9,3,1,3\}$ is equivalent to the former).

In the first case we obtain
$$z_n=-(\alpha n+\beta)+\chi_4(n)+\chi_6(n)+\phi_3(n)$$
$$\zeta_n=3(\alpha n+\beta)+3\alpha-\chi_4(n+1)-\chi_6(n)$$
$$A_n=3(\alpha n+\beta)-\alpha+\chi_4(n+1)+\phi_3(n)-\phi_3(n-1)+\chi_6(n)-\chi_6(n+1)-\chi_6(n-1)$$
$$B_n=\alpha n+\beta+\alpha-\chi_4(n+1)-\phi_3(n+1)+\chi_6(n)+\chi_6(n+1)-\chi_6(n-1)$$
$$C_n=4(\alpha n+\beta)+5\alpha+\chi_4(n)-\chi_4(n+1)-\phi_3(n-1)+\chi_6(n+1)+\chi_6(n-1)$$
$$D_n=\alpha+\chi_4(n)-\chi_4(n+1)+\phi_3(n)-\phi_3(n+1)-\chi_6(n+1)-\chi_6(n-1).$$

Eliminating $x$ from (\ddyo) in this case we obtain for $y$ an equation derived in [\ancil], case 4.4.4, while eliminating $y$ leads for $x$ to case 4.5.1 of the same reference.

In the second case we find
$$z_n=-\chi_6(n)-\chi_6(n-1)$$
$$\zeta_n=2(\alpha n+\beta)+\phi_3(n)+\chi_6(n)$$
$$A_n=4(\alpha n+\beta)-5\alpha+\chi_4(n)-\phi_3(n+1)-\chi_6(n)$$
$$B_n=\alpha-\chi_4(n)+\phi_3(n-1)-\phi_3(n)-\chi_6(n)$$
$$C_n=4(\alpha n+\beta)+\alpha-\chi_4(n)-\phi_3(n+1)-\chi_6(n-1)$$
$$D_n=-\alpha+\chi_4(n)+\phi_3(n)-\phi_3(n-1)-\chi_6(n-1).$$
Eliminating either $x$ or $y$ from (\ddyo) in this case leads to case 4.4.4 of [\ancil].

Since the singularity pattern (7,7,1,1) has already been studied in detail we turn to the case of (7,5,3,1). In this case a pattern of the first class, i.e. $\{A\to C, B\to D, C\to A,D\to B\}$, as explained in the beginning of this section, does exist and in fact with two distinct realisations corresponding to $\{7,5,3,1\}$ and $\{7,3,5,1\}$. 

We find in the first case
$$z_n=-(\alpha n+\beta)+\phi_3(n)+\phi_5(n)$$
$$\zeta_n=3(\alpha n+\beta)+3\alpha/2+\phi_5(n-1)+\phi_5(n+2)$$
$$A_n=5(\alpha n+\beta)/2+\gamma-\phi_3(n+1)+\phi_5(n+1)-\phi_5(n+2)$$
$$B_n=3(\alpha n+\beta)/2-3\alpha-\gamma+\phi_3(n)-\phi_3(n-1)+\phi_5(n)-\phi_5(n-1)+\phi_5(n-2)$$
$$C_n=5(\alpha n+\beta)/2-\gamma-\phi_3(n-1)+\phi_5(n-1)-\phi_5(n-2)$$
$$D_n=3(\alpha n+\beta)/2+3\alpha+\gamma+\phi_3(n)-\phi_3(n+1)+\phi_5(n)-\phi_5(n+1)+\phi_5(n+2).$$
Eliminating either $x$ or $y$ from (\ddyo) in this case leads to case 4.3.4 of [\ancil].

In the second case we obtain
$$z_n=2\chi_6(n)$$
$$\zeta_n=2(\alpha n+\beta)+\phi_3(n)-\chi_6(n)-\chi_6(n+1)$$
$$A_n=3(\alpha n+\beta)+\gamma+\phi_2(n)+\phi_3(n-1)-\phi_3(n)+\chi_6(n)$$
$$B_n=\alpha n+\beta-4\alpha-\gamma-\phi_2(n)-\phi_3(n+1)+\chi_6(n)-2\chi_6(n-1)$$
$$C_n=3(\alpha n+\beta)-3\alpha-\gamma+\phi_2(n)+\phi_3(n)-\phi_3(n-1)+\chi_6(n)$$
$$D_n=\alpha n+\beta+3\alpha+\gamma-\phi_2(n)-\phi_3(n+1)+\chi_6(n)-2\chi_6(n+1).$$

Eliminating either $x$ or $y$ from (\ddyo) in this case leads to case 4.2.5 of [\ancil].

Patterns of what we called the second class do also exist.  

For the pattern $\{A\to C, B\to D, C\to B,D\to A\}$ with (7,5,3,1), three distinct cases do exist corresponding to patterns $\{7,5,3,1\}$, $\{7,3,5,1\}$ and $\{7,1,3,5\}$.

In the first case we have
$$z_n=-(\alpha n+\beta)+5\phi_8(n)$$
$$\zeta_n=3(\alpha n+\beta)+2\alpha+\phi_8(n+6)-\phi_8(n+5)+3\phi_8(n+4)+\phi_8(n+2)-\phi_8(n+1)-2\phi_8(n)$$
$$A_n=2(\alpha n+\beta)-2\alpha-4\phi_8(n+7)+4\phi_8(n+6)+3\phi_8(n+5)+\phi_8(n+3)-\phi_8(n+2)-2\phi_8(n+1)+5\phi_8(n)$$
$$B_n=2(\alpha n+\beta)-4\phi_8(n+6)-\phi_8(n+5)-2\phi_8(n+4)+5\phi_8(n+3)+\phi_8(n+2)+4\phi_8(n+1)+3\phi_8(n)$$
$$C_n=3(\alpha n+\beta)+2\alpha-5(\phi_8(n+5)+\phi_8(n+3)+\phi_8(n+2))$$
$$D_n=\alpha n+\beta+2\alpha-6\phi_8(n+7)-4\phi_8(n+6)-3\phi_8(n+5)-\phi_8(n+3)+\phi_8(n+2)-8\phi_8(n+1).$$

Eliminating $x$ from (\ddyo) in this case we obtain for $y$ equation 4.3.2, derived in [\ancil], while eliminating $y$ leads for $x$ to case 4.4.2 of the same article.

In the second case we find
$$z_n=\alpha+\phi_8(n+1)-\phi_8(n)$$
$$\zeta_n=2(\alpha n+\beta)+\phi_8(n+5)+\phi_8(n)$$
$$A_n=2(\alpha n+\beta)-3\alpha+\phi_8(n+7)-\phi_8(n+6)-\phi_8(n+5)-2\phi_8(n+3)-3\phi_8(n)$$
$$B_n=2(\alpha n+\beta)+\alpha+\phi_8(n+4)+\phi_8(n+3)-\phi_8(n+2)+\phi_8(n+1)$$
$$C_n=4(\alpha n+\beta)+\alpha-\phi_8(n+7)-\phi_8(n+6)+\phi_8(n+5)-2\phi_8(n+4)-\phi_8(n+1)$$
$$D_n=\alpha+\phi_8(n+4)-\phi_8(n+3)-\phi_8(n+2)-2\phi_8(n+1)-\phi_8(n).$$
Eliminating $x$ from (\ddyo) in this case we obtain for $y$ equation 4.2.3 of [\ancil], while eliminating $y$ we find for $x$ equation 4.4.2 of [\ancil].

The third case leads to
$$z_n=\alpha n+\beta+\phi_8(n)$$
$$\zeta_n=\alpha n+\beta+2\alpha+\phi_8(n+7)+\phi_8(n+4)-\phi_8(n+3)+\phi_8(n+2)-\phi_8(n)$$
$$A_n=4(\alpha n+\beta)+\alpha+\phi_8(n+3)+\phi_8(n+1)+\phi_8(n)-\phi_8(n-1)+2\phi_8(n-2)$$
$$B_n=\alpha+\phi_8(n+3)-2\phi_8(n+2)+\phi_8(n+1)+\phi_8(n)-\phi_8(n-1)$$
$$C_n=3(\alpha n+\beta)+5\alpha+2\phi_8(n+4)-\phi_8(n+3)+\phi_8(n+1)+\phi_8(n-1)$$
$$D_n=\alpha n+\beta-\phi_8(n+3)+2\phi_8(n+2)-\phi_8(n+1)+\phi_8(n-1).$$
Eliminating $x$ from (\ddyo) in this case we find for $y$ equation 4.3.2 of [\ancil], while eliminating $y$ we obtain for $x$ equation 4.2.3 of [\ancil].

For the pattern $\{A\to C, B\to D, C\to B, D\to A\}$ with $\{7,3,3,3\}$ (the only possible pattern in this case) we find
$$z_n=\psi_6(n)$$
$$\zeta_n=2(\alpha n+\beta)+\alpha+\phi_3(n)+\tilde\psi_6(n)$$
$$A_n=3(\alpha n+\beta)-2\alpha+\phi_2(n)-\phi_3(n)+\phi_3(n-1)+\psi_6(n+1)+\tilde\psi_6(n+1)$$
$$B_n=\alpha n+\beta-\phi_2(n)+\phi_3(n)+\phi_3(n-1)+\psi_6(n)+\psi_6(n-1)+2\tilde\psi_6(n-1)-\tilde\psi_6(n+1)$$
$$C_n=3(\alpha n+\beta)+2\alpha+\phi_2(n)+\phi_3(n)-\phi_3(n-1)+\psi_6(n-1)-\tilde\psi_6(n+1)$$
$$D_n=\alpha n+\beta-\phi_2(n)+\phi_3(n)+\phi_3(n-1)+\psi_6(n)+\psi_6(n+1)+2\tilde\psi_6(n)+\tilde\psi_6(n+1).$$
Here we have introduced the periodic function $\psi_6(n)$ 
satisfying the relation $\psi_6(n+1)+\psi_6(n-1)=\psi_6(n)$ and which can be expressed in term of the cubic roots of unity, $j,j^2$,  as 
$$\psi_6(n)=\gamma(-j)^n+\delta(-j^2)^n.$$
In the results above $\psi_6$ and $\tilde\psi_6$ are two independent functions of two parameters each. Eliminating either $x$ or $y$ from (\ddyo) in this case leads to case 5.1.2 of [\ancil].

The pattern $\{A\to C, B\to D, C\to B, D\to A\}$ with $\{5,5,5,1\}$, again, yields only one possibility:
$$z_n=\omega_9(n+1)-\omega_9(n)-\omega_9(n-1)+\omega_9(n-2)$$
$$\zeta_n=2(\alpha n+\beta)+\alpha+\omega_9(n)$$
$$A_n=\alpha n+\beta-2\alpha+\phi_2(n)-\omega_9(n+5)-2\omega_9(n+4)+\omega_9(n+3)-\omega_9(n+1)-\omega_9(n)$$
$$B_n=3(\alpha n+\beta)-\phi_2(n)+\omega_9(n+6)+\omega_9(n+5)+\omega_9(n+1)$$
$$C_n=3(\alpha n+\beta)-\phi_2(n)+\omega_9(n+3)+\omega_9(n+2)+\omega_9(n-3)$$
$$D_n=\alpha n+\beta+2\alpha+\phi_2(n)+\omega_9(n+5)-\omega_9(n+4)-\omega_9(n+3)+\omega_9(n+1)-\omega_9(n-1).$$
The periodic function $\omega_9$ obeys the relation $\omega_9(n+3)+\omega_9(n-3)+\omega_9(n)=0$ and can be expressed in terms of 6 ninth roots of unity as $$\omega_9(n)=\sum_{\ell=1,2,4,5,7,8}c_{\ell}\exp(2i\pi\ell n/9).$$ Eliminating either $x$ or $y$ from (\ddyo) in this case leads to case 4.2.2 of [\ancil].
 
The last pattern we must examine is (5,5,3,3). Clearly a first class, i.e. pattern $\{A\to C, B\to D, C\to A, D\to B\}$ with steps  $\{5,3,5,3\}$ is just an artificial doubling of the case IV of [\second]. Eliminating either of the two variables leads to an equation already derived in [\ancil], equation 5.2.7. As shown in [\sustem] this equation is a double-step evolution obtained from case IV. 

However, another pattern exists for pattern (5,5,3,3) giving rise to a discrete Painlev\'e equation. 

For the pattern $\{A\to C, B\to D, C\to A, D\to B\}$ with $\{5,5,3,3\}$ we find
$$z_n=\chi_6(n)+\chi_6(n-1)$$
$$\zeta_n=2(\alpha n+\beta)-\chi_6(n)$$
$$A_n=2(\alpha n+\beta)+\gamma+\phi_4(n)+\chi_6(n)$$
$$B_n=2(\alpha n+\beta)-4\alpha-\gamma-\phi_4(n)+\chi_6(n)$$
$$C_n=2(\alpha n+\beta)-2\alpha-\gamma-\phi_4(n+2)+\chi_6(n-1)$$
$$D_n=2(\alpha n+\beta)+2\alpha+\gamma+\phi_4(n+2)+\chi_6(n-1).$$
Eliminating either $x$ or $y$ from (\ddyo) in this case leads to case 4.1 of [\ancil] after some recombination of the parameters.

Finally we have two cases with a pattern $\{A\to C, B\to D, C\to B, D\to A\}$ and steps $\{5,5,3,3\}$ and $\{5,3,5,3\}$ respectively.

In the first case we find
$$z_n=\chi_6(n)+\chi_6(n-1)$$
$$\zeta_n=2(\alpha n+\beta)-\chi_6(n)$$
$$A_n=2(\alpha n+\beta)-2\alpha+\chi_8(n)+\chi_6(n)$$
$$B_n=2(\alpha n+\beta)-2\alpha-\chi_8(n)+\chi_6(n)$$
$$C_n=2(\alpha n+\beta)+\chi_8(n+2)+\chi_6(n-1)$$
$$D_n=2(\alpha n+\beta)-\chi_8(n+2)+\chi_6(n-1).$$
Eliminating either $x$ or $y$ from (\ddyo) in this case leads again to case 4.1 of [\ancil].

In the second case we obtain
$$z_n=\alpha n+\beta+\chi_6(n)+\chi_8(n)$$
$$\zeta_n=\alpha n+\beta-\chi_6(n+1)-\chi_8(n)$$
$$A_n=2(\alpha n+\beta)-4\alpha+\chi_8(n)-\chi_8(n-2)-\chi_8(n+1)-\chi_8(n-1)+\chi_6(n)$$
$$B_n=2(\alpha n+\beta)+\chi_8(n)+\chi_8(n-2)+\chi_8(n+1)-\chi_8(n-1)-\chi_6(n)$$
$$C_n=3(\alpha n+\beta)+\chi_8(n-1)-\chi_8(n+1)+\chi_8(n+2)-2\chi_6(n-1)$$
$$D_n=\alpha n+\beta+\chi_8(n+1)-\chi_8(n-1)-\chi_8(n+2)-2\chi_6(n-1).$$
Eliminating $x$ from (\ddyo) in this case we obtain for $y$ equation 5.2.4 of [\ancil], while eliminating $y$ leads for $x$ to case 4.1 of the same article.

\bigskip
4. {\scap Singularity analysis of the even and odd steps case}
\medskip

The final set we shall analyse is when two of the singularity steps $M,N,P,Q$ are even and some are odd. Without loss of generality we can choose to enter the singularity through $A$ and exit it through $C$ in an odd number of steps. As a consequence if we enter the singularity through $B$ we must exit it in an even number of steps. Two possibilities arise: either this singularity exits through $B$ or it exits through $A$. In the latter case we have necessarily $N=4$. In the first case we can choose the singularity entering though $C$ to exit exit through $A$. In the latter case two branches do exist: either $C$ exits through $B$ or it exits through $D$, in which case $P=4$. Putting all this together we have three distinct classes of patterns:
(i) $\{A\to C, B\to B, C\to A, D\to D\}$, (ii) $\{A\to C, B\to A, C\to B,D\to D\}$ and (iii) $\{A\to C, B\to A, C\to D, D\to B\}$. Class (i) comprises all possible even and odd steps collections, class (ii) consists of patterns with at least one step of length 4, while in class (iii) two length-4 steps must exist.

The sets of patterns which are possible for class (i) are: (12,2,1,1), (11,2,2,1), (10,4,1,1), (10,3,2,1), (9,4,2,1), (9,3,2,2), (8,6,1,1), (8,5,2,1), (8,4,3,1), (8,3,3,2), (7,6,2,1), (7,5,2,2), (7,4,4,1), (7,4,3,2), (6,6,3,1), (6,5,4,1), (6,5,3,2), (6,4,3,3), (5,5,4,2), (5,4,4,3). Nine of those sets do also exist for class (ii) while class (iii) comprises just two sets: (7,4,4,1) and (5,4,4,3). Note that the allowed permutations of $(M,N,P,Q)$ do not lead to any new results since the pattern is identical to that initially considered, up to a renaming of $(A,B,C,D)$ and, in some cases, a reversal of the evolution direction. 
We shall start with the singularity analysis of patterns of class (i). The cases  (12,2,1,1), (10,4,1,1), and (8,6,1,1) are already known ones, first obtained in [\first], corresponding to cases VII, VIII-IX and VI of [\second]. We need not repeat the results for them here. 

Class (i), with singularity pattern $\{A\to C, B\to B, C\to A, D\to D\}$.

Steps $\{11,2,1,2\}$
$$z_n=-4(\alpha n+\beta)+\phi_3(n)+\phi_5(n)$$
$$\zeta_n=6(\alpha n+\beta)+3\alpha+\phi_5(n-2)$$
$$A_n=3(\alpha n+\beta)-7\alpha+\phi_2(n)+\phi_3(n)-\phi_3(n-1)-\phi_5(n-1)-\phi_5(n-2)$$
$$B_n=\alpha n+\beta+\alpha-\phi_2(n)- \phi_3(n + 1) + \phi_5(n + 2) - \phi_5(n + 1) + \phi_5(n)$$
$$C_n=3(\alpha n+\beta)+7\alpha+\phi_2(n)-\phi_3(n + 1) + \phi_3(n) -\phi_5(n + 2) - \phi_5(n + 1)$$
$$D_n=\alpha n+\beta-\alpha-\phi_2(n)- \phi_3(n - 1)+ \phi_5(n - 2) - \phi_5(n - 1) + \phi_5(n).$$
Eliminating $y$ of $x$ we obtain for $x$ and $y$ respectively an equation which is precisely 3.1 of [\ancil].

Steps $\{3,10,1,2\}$
$$z_n=2\alpha+\phi_7(n + 3) - \phi_7(n + 1)$$
$$\zeta_n=2(\alpha n+\beta)+\phi_7(n)+\phi_7(n+1)$$
$$A_n=-(\alpha n+\beta)-5\alpha/2+\phi_2(n)+\phi_7(n)-\phi_7(n+1)-\phi_7(n+2)$$
$$B_n=5(\alpha n+\beta)+5\alpha/2-\phi_2(n)+2\phi_7(n - 1) + 2\phi_7(n + 3) + \phi_7(n + 2) - \phi_7(n + 1) + \phi_7(n)$$
$$C_n=3(\alpha n + \beta)+ 7\alpha/2+\phi_2(n)+\phi_7(n + 3) + \phi_7(n + 2) + \phi_7(n)$$
$$D_n=\alpha n + \beta+\alpha/2-\phi_2(n)+\phi_7(n + 3) - \phi_7(n + 2) + \phi_7(n).$$
Eliminating $x$ in this case we obtain for $y$ equation 4.4.3 of [\ancil], while eliminating $y$ leads for $x$ to case 3.3 of the same article.

Steps $\{9,4,1,2\}$
$$z_n=-3(\alpha n+\beta)+2\alpha+\phi_4(n)+\phi_5(n)$$
$$\zeta_n=5(\alpha n+\beta)+\phi_4(n+2)$$
$$A_n=2(\alpha n+\beta)-7\alpha+\phi_4(n + 1) + \phi_4(n)- \phi_5(n - 1) + \phi_5(n + 2) + \phi_5(n)$$
$$B_n=2(\alpha n+\beta)+\alpha+\phi_4(n + 1) + \phi_4(n)+\phi_5(n - 1) - \phi_5(n + 2) + \phi_5(n)$$
$$C_n=3(\alpha n+\beta)+5\alpha-\phi_4(n+1)- \phi_5(n + 3) - \phi_5(n + 1) + \phi_5(n)$$
$$D_n=\alpha n+\beta-\alpha-\phi_4(n + 1)-2\phi_4(n - 1)-\phi_5(n+2)-\phi_5(n-1).$$
Eliminating $y$ we obtain for $x$ equation 3.2 of [\ancil], while eliminating $x$ leads for $y$ to equation 4.5.2.

Steps $\{9,2,3,2\}$
$$z_n=-2(\alpha n+\beta)+\phi_7(n)+\phi_7(n+1)$$
$$\zeta_n=4(\alpha n+\beta)+2\alpha+\phi_7(n-3)+\phi_7(n-2)+\phi_7(n+1)$$
$$A_n=3(\alpha n+\beta)-9\alpha/2+\phi_2(n)-2\phi_7(n - 2)-\phi_7(n - 1)-\phi_7(n + 2)+\phi_7(n)$$
$$B_n=\alpha n+\beta+\alpha/2-\phi_2(n)-\phi_7(n - 1)+\phi_7(n)-\phi_7(n+2)$$
$$C_n=3(\alpha n+\beta)+9\alpha/2-\phi_2(n)-\phi_7(n - 1)+\phi_7(n + 1)-\phi_7(n +2)-2\phi_7(n +3)$$
$$D_n=\alpha n+\beta-al/2+\phi_2(n)- \phi_7(n - 1)- \phi_7(n + 2) + \phi_7(n + 1).$$
Here eliminating either $y$ of $x$ leads, for the other variable to an equation which is 4.4.3 of [\ancil].

Steps $\{5,8,1,2\}$
$$z_n=-\alpha n-\beta+\phi_3(n)+\phi_2(n)+\chi_8(n)$$
$$\zeta_n=3(\alpha n+\beta)+6\alpha+\phi_2(n)-\chi_8(n)$$
$$A_n=-4\alpha- \phi_3(n - 1) + \phi_3(n)- \chi_8(n - 1) + \chi_8(n + 2) - \chi_8(n + 1) + \chi_8(n)$$
$$B_n=4(\alpha n + \beta) + 10\alpha-\phi_3(n+1)- \chi_8(n - 1) - \chi_8(n + 2) + \chi_8(n + 1) + \chi_8(n)$$
$$C_n=3(\alpha n + \beta) + 10\alpha+\phi_2(n)- \phi_3(n + 1) + \phi_3(n)- \chi_8(n + 3) + \chi_8(n + 2) - \chi_8(n + 1)$$
$$D_n=\alpha n + \beta+ 2\alpha+3\phi_2(n)- \phi_3(n - 1)+\chi_8(n + 3) - \chi_8(n + 2) + \chi_8(n + 1).$$
Here we obtain for $x$ a trihomographic equation which is 3.4 of [\ancil] and for $y$ equation 4.3.3 of the same article.

Steps $\{3,8,1,4\}$
$$z_n=\alpha+\phi_8(n)-\phi_8(n+1)$$
$$\zeta_n=2(\alpha n+\beta)-\phi_8(n)-\phi_8(n-3)$$
$$A_n= - 2\alpha - \phi_8(n - 3) + \phi_8(n - 2) - \phi_8(n - 1) + \phi_8(n)$$
$$B_n=4(\alpha n + \beta)- 2\phi_8(n - 4) + \phi_8(n - 3) - \phi_8(n - 2) - \phi_8(n - 1) - 2\phi_8(n + 1) + \phi_8(n)$$
$$C_n=2(\alpha n  +\beta) + 2\alpha- \phi_8(n - 3) + \phi_8(n - 2) - \phi_8(n - 1) - \phi_8(n + 1)$$
$$D_n=2(\alpha n + \beta)- \phi_8(n - 3) - \phi_8(n - 2) + \phi_8(n - 1) - \phi_8(n + 1).$$

Eliminating $y$ we obtain for $x$ equation 4.4.2 of [\ancil], while eliminating $x$ leads for $y$ to equation 4.2.3.

Steps $\{3,8,3,2\}$
$$z_n=\alpha n + \beta + 4\alpha+\phi_8(n + 4)$$
$$\zeta_n=\alpha n+\beta+\phi_8(n)$$
$$A_n=\phi_8(n + 3) - \phi_8(n + 2) - \phi_8(n + 1) + \phi_8(n)$$
$$B_n=4(\alpha n+\beta) + 6\alpha+2\phi_8(n - 1) + 2\phi_8(n + 4) - \phi_8(n + 3) + \phi_8(n + 2) + \phi_8(n + 1) - \phi_8(n)$$
$$C_n=3(\alpha n+\beta) + 6\alpha+\phi_8(n + 4) + \phi_8(n + 3) - \phi_8(n + 2) + \phi_8(n + 1) + \phi_8(n) $$
$$D_n=\alpha n+\beta + 2\alpha +\phi_8(n + 4) - \phi_8(n + 3) + \phi_8(n + 2) - \phi_8(n + 1) + \phi_8(n).$$
 Again, the equation for $x$ turns out to be 4.4.2 of [\ancil] but now the equation for $y$ is 5.1.3.
 
Steps $\{7,6,1,2\}$
$$z_n=-2(\alpha n+\beta)+2\phi_7(n)+2\phi_7(n+1)$$
$$\zeta_n=4(\alpha n+\beta)+4\alpha-\phi_7(n+1)-\phi_7(n)+\phi_7(n+3)$$
$$A_n=\alpha n+\beta-9\alpha/2+\phi_2(n)+\phi_7(n - 3)-3\phi_7(n - 1) + \phi_7(n + 1)-\phi_7(n)$$
$$B_n= 3(\alpha n +\beta)+ 9\alpha/2- \phi_2(n) - \phi_7(n - 3) + \phi_7(n - 1) + 2\phi_7(n + 2) + 3\phi_7(n + 1) + 3\phi_7(n)$$
$$C_n=3(\alpha n+\beta) + 15\alpha/2+\phi_2(n) - \phi_7(n - 2) + \phi_7(n + 3) - \phi_7(n + 2) + 2\phi_7(n)$$
$$D_n=\alpha n + \beta+\alpha/2-\phi_2(n) - \phi_7(n - 1) - \phi_7(n -3) + \phi_7(n + 1) - \phi_7(n).$$
Eliminating either $y$ of $x$ leads, for the other variable to an equation which is 4.4.3 of [\ancil].

Steps $\{7,2,5,2\}$
$$z_n=- \omega_6(n -2)- \omega_6(n+1)$$
$$\zeta_n=2(\alpha n+\beta)+\phi_3(n)+\omega_6(n)$$
$$A_n=3(\alpha n+\beta)-7\alpha/2+\phi_2(n)+\phi_3(n-1)-\phi_3(n)-\omega_6(n+1)$$
$$B_n=\alpha n + \beta-\alpha/2-\phi_2(n)- \phi_3(n + 1) - 2\omega_6(n - 2) + \omega_6(n - 1) + \omega_6(n - 3)$$
$$C_n=3(\alpha n+\beta)+\alpha/2+\phi_2(n) - \phi_3(n - 1) + \phi_3(n) - \omega_6(n - 2)$$
$$D_n=\alpha n + \beta-\alpha/2-\phi_2(n) - \phi_3(n + 1) + \omega_6(n + 2) - 2\omega_6(n + 1) + \omega_6(n).$$
The periodic function $\omega_6$ obeys the relation $\omega_6(n+2)+\omega_6(n-2)-\omega_6(n)=0$ and can be expressed in terms of 6 ninth roots of unity as $$\omega_6(n)=\sum_{\ell=1,5,7,11}c_{\ell}\exp(i\pi\ell n/6).$$ 
Eliminating either of the two variables leads to the same equation, namely 4.2.4 of [\ancil].

Steps $\{7,4,1,4\}$
$$z_n=-2(\alpha n+\beta)+\alpha+2\phi_6(n)+2\phi_6(n+2)$$
$$\zeta_n=4(\alpha n+\beta)-\phi_6(n+3)-\phi_6(n)$$
$$A_n=2(\alpha n+\beta)-6\alpha+\chi_4(n)- 2\phi_6(n - 1) + \phi_6(n -2)  + 2\phi_6(n + 2) - \phi_6(n + 1) + 2\phi_6(n)$$
$$B_n=2(\alpha n + \beta)- \chi_4(n) - \phi_6(n - 2) + \phi_6(n + 1) + 2\phi_6(n)$$
$$C_n=2(\alpha n +\beta)+ 4\alpha- \chi_4(n + 1) - 2\phi_6(n - 1) - \phi_6(n -2) - 4\phi_6(n + 3) - 3\phi_6(n + 1)$$
$$D_n=2(\alpha n+\beta) - 2\alpha+\chi_4(n + 1) - \phi_6(n -2) + 2\phi_6(n + 2) + \phi_6(n + 1).$$
Here, also, elimination results to the same equation, in this case 4.4.4 of [\ancil].

Steps $\{7,4,3,2\}$
$$z_n=-\alpha n-\beta+\phi_8(n)$$
$$\zeta_n=3(\alpha n +\beta) + 4\alpha - \phi_8(n - 1) - \phi_8(n + 4) - \phi_8(n + 1)$$
$$A_n=2(\alpha n+\beta) -\alpha- 2\phi_8(n - 2) + \phi_8(n - 1) - \phi_8(n + 3) + \phi_8(n + 2) - \phi_8(n + 1)$$
$$B_n=2(\alpha n+\beta) +3\alpha- \phi_8(n - 1) - \phi_8(n + 3) - \phi_8(n + 2) + \phi_8(n + 1)$$
$$C_n=3(\alpha n+\beta) +7\alpha- \phi_8(n - 1) - 2\phi_8(n + 4) + \phi_8(n + 3) - \phi_8(n + 2) - \phi_8(n + 1) + \phi_8(n)$$
$$D_n=\alpha n+\beta+\alpha- \phi_8(n - 1) - \phi_8(n + 3) + \phi_8(n + 2) - \phi_8(n + 1) + \phi_8(n).$$
The equation for $x$ obtained by elimination turns out to be 4.2.3 of [\ancil], while for $y$ we find 4.1.3.

Steps $\{3,6,1,6\}$
$$z_n=\omega_9(n - 1) - \omega_9(n + 2)$$
$$\zeta_n=2(\alpha n+\beta)+\alpha+\omega_9(n+2)-\omega_9 n$$
$$A_n=\alpha n+\beta-2\alpha+\phi_2(n)+2\omega_9(n+1)+\omega_9(n-1)$$
$$B_n=3(\alpha n+\beta)-\phi_2(n)+\omega_9(n-4) - \omega_9(n + 2)$$
$$C_n=\alpha n+\beta+2\alpha+\phi_2(n)- \omega_9(n + 2) - 2\omega_9(n)$$
$$D_n=3(\alpha n+\beta)-\phi_2(n)+\omega_9(n - 1) - \omega_9(n -4).$$
Both equations, either for $x$ or for $y$, are equation 4.2.2 of [\ancil].

Steps $\{5,6,1,4\}$
$$z_n=-\alpha n-\beta+\alpha - \phi_8(n - 1) - \phi_8(n + 2) - \phi_8(n)$$
$$\zeta_n=3(\alpha n+\beta)+\phi_8(n)$$
$$A_n=\alpha n+\beta - 4\alpha- \phi_8(n - 3) - \phi_8(n + 3) - 2\phi_8(n + 2) - \phi_8(n)$$
$$B_n=3(\alpha n + \beta)+\phi_8(n - 3) + \phi_8(n + 3) - \phi_8(n)$$
$$C_n=2(\alpha n +\beta)+ 3\alpha- \phi_8(n - 3) - \phi_8(n - 1) + \phi_8(n + 3) - \phi_8(n + 2)$$
$$D_n=2(\alpha n +\beta)-\alpha+\phi_8(n - 3) - \phi_8(n - 1) - \phi_8(n + 3) - \phi_8(n + 2).$$
Eliminating $y$ we obtain for $x$ equation 4.3.2 of [\ancil], while eliminating $x$ leads for $y$ to equation 4.2.3.

Steps $\{5,6,3,2\}$
$$z_n=2\alpha+\omega_9(n+1)-\omega_9(n-1)$$
$$\zeta_n=2(\alpha n+\beta)-\omega_9(n+2)$$
$$A_n=\alpha n+\beta-3\alpha/2+\phi_2(n)- \omega_9(n - 2) - 2\omega_9(n - 1) - \omega_9(n + 3) - \omega_9(n + 2)$$
$$B_n=3(\alpha n+\beta)+3\alpha/2-\phi_2(n)+\omega_9(n - 2) + \omega_9(n + 3) + \omega_9(n + 2)$$
$$C_n=3(\alpha n+\beta)+7\alpha/2-\phi_2(n+\omega_9(n + 5) + \omega_9(n + 4) - \omega_9(n + 3) + 2\omega_9(n + 1))$$
$$D_n=\alpha n+\beta+\alpha/2+\phi_2(n)- 2\omega_9(n - 1) - \omega_9(n + 5) - \omega_9(n + 4) + \omega_9(n + 3) - 2\omega_9(n + 2).$$
After elimination of $y$ we find for $x$ equation 4.2.2 of [\ancil], and similarly equation 5.2.8 for $y$.

Steps $\{3,6,3,4\}$
$$z_n=\alpha n+\beta+\chi_4(n)+\chi_{10}(n)$$
$$\zeta_n=\alpha n+\beta -\alpha+\chi_{10}(n - 1) + \chi_{10}(n + 4) - \chi_{10}(n) + \chi_4(n - 1)$$
$$A_n=\alpha n +\beta-\alpha-\chi_{10}(n - 2) - \chi_{10}(n - 1)+ \chi_{10}(n)  - \chi_{10}(n + 2) + \chi_{10}(n + 1)+\chi_4(n - 1)$$
$$B_n=3(\alpha n +\beta)-3\alpha+\chi_{10}(n - 2) - \chi_{10}(n - 1) + \chi_{10}(n + 2) - \chi_{10}(n + 1) + \chi_{10}(n) + 2\chi_4(n - 2) - \chi_4(n - 1) + 2\chi_4(n)$$
$$C_n=2(\alpha n +\beta)-\alpha-\chi_{10}(n-1)- \chi_{10}(n + 3) - \chi_{10}(n + 2) + \chi_{10}(n + 1) + \chi_4(n - 1) + \chi_4(n)$$
$$D_n=2(\alpha n +\beta)-\alpha+\chi_{10}(n - 1) + \chi_{10}(n -2) + \chi_{10}(n + 2) - \chi_{10}(n + 1) + \chi_4(n - 1) + \chi_4(n).$$
The elimination of $y$ leads for $x$ to equation 5.1.1 of [\ancil], while elimination of $x$ gives equation 5.2.3 for $y$.

Steps $\{5,4,5,2\}$
$$z_n=\alpha n+\beta+\phi_2(n)+\chi_{12}(n)$$
$$\zeta_n=\alpha n+\beta-2\alpha+\phi_2(n)-\chi_{12}(n)$$
$$A_n=2(\alpha n +\beta)-3\alpha+\chi_{12}(n - 2) - \chi_{12}(n - 1) + \chi_{12}(n + 3) + \chi_{12}(n + 2) - \chi_{12}(n + 1) + \chi_{12}(n)$$
$$B_n=2(\alpha n +\beta)-3\alpha- \chi_{12}(n - 2) - \chi_{12}(n - 1) - \chi_{12}(n + 3) - \chi_{12}(n + 2) + \chi_{12}(n + 1) + \chi_{12}(n)$$
$$C_n=3(\alpha n +\beta)-3\alpha + \phi_2(n)- \chi_{12}(n + 5) + \chi_{12}(n + 4) - \chi_{12}(n + 3) + \chi_{12}(n + 2) - \chi_{12}(n + 1)$$
$$D_n=\alpha n +\beta-\alpha + 3\phi_2(n)+\chi_{12}(n + 5) - \chi_{12}(n + 4) + \chi_{12}(n + 3) - \chi_{12}(n + 2) + \chi_{12}(n + 1) + 3\phi_2(n).$$
Here the equation we obtain for $x$ is 4.1 of[\ancil] while that for $y$ is 5.2.6.

Steps $\{5,4,3,4\}$
$$z_n=-\chi_{10}(n-1)-\chi_{10}(n)$$
$$\zeta_n=2(\alpha n+\beta)+\chi_{10}(n)$$
$$A_n=2(\alpha n+\beta)-3\alpha-\chi_{10}(n+2)-\chi_{10}(n)+\chi_4(n)$$
$$B_n=2(\alpha n +\beta)-\alpha+\chi_{10}(n + 2) - \chi_{10}(n)-\chi_4(n)$$
$$C_n=2(\alpha n +\beta)+\alpha-\chi_{10}(n -1) + \chi_{10}(n + 2)+\chi_4(n)$$
$$D_n=2(\alpha n +\beta)-\alpha- \chi_{10}(n - 1) - \chi_{10}(n + 2) -\chi_4(n).$$
Both equations obtained by elimination of one variable are equation 5.2.3 of [\ancil].

Class (ii), with singularity pattern $\{A\to C, B\to A, C\to B,D\to D\}$.

The case with steps (10,4,1,1), belonging to class (ii), was already obtained in [\second], case IX there. 

Steps $\{9,4,1,2\}$
$$z_n=-\alpha n-\beta+\phi_4 n+\phi_3(n)$$
$$\zeta_n=3(\alpha n+\beta)+2\gamma+\phi_4(n+2)$$
$$A_n=4(\alpha n+\beta)-5\alpha+5\gamma+\phi_2(n) - \phi_3(n +1) - \phi_4(n + 2) + \phi_4(n + 1)$$
$$B_n=-\alpha-\gamma- \phi_2(n) - \phi_3(n - 1) + \phi_3(n) - \phi_4(n - 1) + \phi_4(n)$$
$$C_n=3(\alpha n+\beta)+\alpha+3\gamma-\phi_2(n) + \phi_3(n) - \phi_3(n + 1) - \phi_4(n - 1) - 2\phi_4(n + 1)$$
$$D_n=\alpha n +\beta-\alpha+\gamma+\phi_2(n) - \phi_3(n - 1) - \phi_4(n - 1).$$
The equation obtained for $x$, after elimination of $y$, is 3.4 of [\ancil], while that for $y$ is 4.5.1 of the same paper.

Steps $\{3,4,1,8\}$
$$z_n=(\alpha n+\beta)/2+\phi_5(n)+\phi_3(n)$$
$$\zeta_n=3(\alpha n+\beta)/2+9\alpha/2+2\gamma-\phi_5(n+1)-\phi_5(n)$$
$$A_n=5(\alpha n+\beta)/2+4\alpha+2\gamma- \phi_3(n + 1) + \phi_5(n + 2) - \phi_5(n -2)$$
$$B_n=3(\alpha n+\beta)/2+2\alpha+2\gamma-\phi_5(n + 2) - \phi_5(n - 1)$$
$$C_n=\alpha+\phi_3(n) - \phi_3(n + 1) + \phi_5(n -2) - \phi_5(n + 1)$$
$$D_n=4(\alpha n+\beta)+8\alpha+4\gamma-\phi_3(n -1)-\phi_5(n -2)- \phi_5(n + 1).$$
Eliminating $y$ we obtain for $x$ equation 4.4.1 of [\ancil], while eliminating $x$ leads for $y$ to equation 4.3.4.

Steps $\{7,4,1,4\}$
$$z_n=-(\alpha n+\beta)/2+\phi_5 n+\phi_3 n$$
$$\zeta_n=5(\alpha n+\beta)/2-5\alpha/2+\gamma+\phi_3(n - 1)$$
$$A_n=7(\alpha n+\beta)/2-8\alpha+2\gamma-\phi_3(n-1) - \phi_5(n + 1) - \phi_5(n - 2)$$
$$B_n=(\alpha n+\beta)/2-2\alpha- \phi_3(n + 2) - \phi_5(n - 1) - \phi_5(n + 2) + \phi_5(n)$$
$$C_n=2(\alpha n+\beta)-\alpha+\gamma-\phi_3(n + 1) + \phi_5(n + 3) - \phi_5(n + 1) + \phi_5(n)$$
$$D_n=2(\alpha n+\beta)-4\alpha+\gamma-\phi_3(n + 1)- \phi_5(n + 3) + \phi_5(n + 1) + \phi_5(n).$$
Elimination leads to precisely the same equations as in the previous case.

Steps $\{7,4,3,2\}$
$$z_n=\alpha+2\gamma-\phi_6(n - 1)- \phi_6(n + 1) - \phi_6(n)$$
$$\zeta_n=2(\alpha n+\beta)+\phi_6(n)$$
$$A_n=3(\alpha n+\beta) - \alpha+ 5\gamma+\phi_2( n)+\phi_6(n - 1) - \phi_6(n)$$
$$B_n=\alpha n+\beta-\alpha-\gamma- \phi_2(n)- \phi_6(n - 1) - 2\phi_6(n + 1) - \phi_6(n)$$
$$C_n=3(\alpha n+\beta)+2\alpha+3\gamma+ \phi_2(n )+2\phi_6(n + 3) + \phi_6(n + 1)$$
$$D_n=\alpha n+\beta+\gamma- \phi_2(n)- 2\phi_6(n - 1) - 2\phi_6(n + 3) - 3\phi_6(n + 1).$$
We find the $x$ equation 4.2.5 of [\ancil] while for $y$ we obtain 5.1.2 for $y$.

Steps $\{5,4,1,6\}$ 
$$z_n=4\alpha+\phi_2(n)+\phi_3(n+1)-\phi_3(n)-\psi_6(n-1)$$
$$\zeta_n=2(\alpha n+\beta)+\alpha+\phi_2(n)+\phi_3(n)+\psi_6(n)$$
$$A_n=3(\alpha n+\beta)+10\alpha+\tilde\phi_2(n)+\phi_3(n+1)-\phi_3(n) + \psi_6(n-1)$$
$$B_n=\alpha n+\beta-4\alpha-\tilde\phi_2(n)+\phi_3(n-1)-2\phi_3(n)- \psi_6(n-1)$$
$$C_n=\alpha n+\beta+4\alpha+\phi_2(n)-\tilde\phi_2(n)+2\phi_3(n+1)+2\psi_6(n)$$
$$D_n=3(\alpha n+\beta)+6\alpha+3\phi_2(n)+\tilde\phi_2(n)- 2\psi_6(n - 1).$$
Note that the combination $\phi_2(n)+\phi_3(n)+\psi_6(n)$ appearing in $\zeta_n$ could be combined to a single $\phi_6(n)$ function, but then this parametrisation would have been quite inconvenient for the remaining veriables.
Elimination of either of the dependent variables leads to equation 4.2.5 of [\ancil] for the other one.

Steps $\{3,4,3,6\}$
$$z_n=\alpha n+\beta+\chi_6(n)+\chi_4(n)$$
$$\zeta_n=\alpha n+\beta-2\alpha+2\gamma+\chi_6(n-2)+\chi_4(n-2)$$
$$A_n=2(\alpha n+\beta)-\alpha+\gamma+\phi_2(n)+\chi_4(n + 1) + \chi_4(n) -\chi_6(n )$$
$$B_n=2(\alpha n+\beta)-5\alpha+3\gamma+\phi_2(n)+\chi_4(n +1) + \chi_4(n) + \chi_6(n)$$
$$C_n=\alpha n+\beta-\alpha+\gamma+\phi_2(n)+\chi_4(n+1)-2\chi_6(n+2).$$
$$D_n=3(\alpha n+\beta)-3\alpha+3\gamma-\phi_2(n)- \chi_4(n + 1) + 2\chi_6(n - 2) + 2\chi_6(n + 2) + 2\chi_6(n).$$
When we eliminate $x$ we find for $y$ equation 5.1.1 of [\ancil]. Similarly eliminating $y$ we find for $x$ equation 5.2.10 (which, in fact, is identical to the case 5.2.9 of the same paper).

Steps $\{5,4,5,2\}$
$$z_n=\alpha n+\beta+\phi_2(n)+\chi_8(n)+\chi_8(n+1)$$
$$\zeta_n=\alpha n+\beta+2\gamma-\chi_8(n)  - \chi_8(n + 1)$$
$$A_n=2(\alpha n+\beta)-2\alpha+\gamma+\phi_2(n)-2\chi_8(n-1)$$
$$B_n=2(\alpha n+\beta)+3\gamma-\phi_2(n)+2\chi_8(n + 1)$$
$$C_n=3(\alpha n+\beta)+3\gamma-\phi_2(n)+\chi_8(n)  - \chi_8(n + 1)$$
$$D_n=\alpha n+\beta+\gamma+4\phi_2(n)- \chi_8(n) + \chi_8(n + 1).$$
Eliminating $y$ we find for $x$ equation 4.1 of [\ancil], while eliminating $x$ we obtain for $y$ to equation 5.2.4.

Steps $\{5,4,3,4\}$
$$z_n=(\alpha n+\beta)/2+\phi_7(n)$$
$$\zeta_n=3(\alpha n+\beta)/2+3\alpha/2+\gamma+\phi_7(n+4)+\phi_7(n+2)+\phi_7(n-1)$$
$$A_n=5(\alpha n +\beta)/2+\gamma+2\phi_7(n - 2) - \phi_7(n - 1) + \phi_7(n + 3) + \phi_7(n + 2) + 2\phi_7(n)$$
$$B_n=3(\alpha n +\beta)/2+\gamma+\phi_7(n - 1) + \phi_7(n + 3) - \phi_7(n + 2) + 2\phi_7(n + 1)$$
$$C_n=2(\alpha n + \beta)+2\alpha+\gamma+\phi_7(n - 1) + 2\phi_7(n + 4) - \phi_7(n + 3) + \phi_7(n + 2) + \phi_7(n)$$
$$D_n=2(\alpha n + \beta)+\alpha+\gamma+\phi_7(n - 1) + \phi_7(n + 3) + \phi_7(n + 2) + \phi_7(n).$$
Elimination of $x$ leads for $y$ to equation 5.2.7 of [\ancil], while when we eliminate $y$ we find equation 5.2.2 for $x$.

Class (iii), with singularity pattern $\{A\to C, B\to A, C\to D, D\to B\}$.

Steps $\{7,4,4,1\}$
$$z_n=\gamma+\phi_3(n)$$
$$\zeta_n=2(\alpha n+\beta)+\tilde{\phi_3}(n)$$
$$A_n=3(\alpha n +\beta)+ \delta+\phi_2(n) - \phi_3(n + 1) + \tilde{\phi_3}(n - 1) - \tilde{\phi_3}(n)$$
$$B_n=\alpha n+\beta-4\alpha+2\gamma-\delta-\phi_2(n)- \phi_3(n - 1) + \phi_3(n) + \tilde{\phi_3}(n - 1) + \tilde{\phi_3}(n)$$
$$C_n=3(\alpha n +\beta)-3\alpha+4\gamma-\delta+\phi_2(n)- \phi_3(n + 2)- \tilde{\phi_3}(n - 1) + \tilde{\phi_3}(n)$$
$$D_n=\alpha n +\beta+3\alpha-2\gamma+\delta-\phi_2(n)+ \phi_3(n) - \phi_3(n + 1) - \tilde{\phi_3}(n + 1).$$
Elimination of either $x$ or $y$ leads to equation 4.3.1 of [\ancil].

Steps $\{5,4,4,3\}$
$$z_n=2(\alpha n+\beta)/3+\gamma-\phi_5(n-1)-\phi_5(n)$$
$$\zeta_n=4(\alpha n+\beta)/3+\phi_5(n)$$
$$A_n=7(\alpha n+\beta)/3-4\alpha/3+\delta+\phi_2(n)-\phi_5(n-3)-\phi_5(n)$$
$$B_n=5(\alpha n + \beta)/3-4\alpha/3+2\gamma-\delta-\phi_2(n)+\phi_5(n - 3) - \phi_5(n)$$
$$C_n=7(\alpha n + \beta)/3+3\gamma-\delta-\phi_2(n)- \phi_5(n - 1) - \phi_5(n + 2)$$
$$D_n=5(\alpha n + \beta)/3-\gamma+\delta+\phi_2(n)- \phi_5(n - 1) + \phi_5(n + 2).$$
Again the  same equation is obtained by elimination of either of the two variables, namely 5.2.5 of [\ancil].

\bigskip
{\scap 5. Conclusion and outlook}
\medskip

In [\first] we obtained a collection of discrete Painlev\'e equations with E$_8^{(1)}$ symmetry which can be expressed in trihomographic form, either symmetric or asymmetric. The method used there was the one we call deautonomisation, namely: starting from an autonomous mapping we allow the various parameters to be functions of the independent variable and determine their precise form through the application of an integrability criterion. Five  symmetric and seven asymmetric cases were thus identified. Feeling that the existence of an autonomous starting point was a rather constraining assumption, we asked ourselves whether there might exist more discrete Painlev\'e equations with E$_8^{(1)}$ symmetry, cast into an asymmetric trihomographic form, and which can be obtained by a direct application of the singularity confinement criterion without that assumption. The result was surprising even to us. While we expected another dozen cases to emerge we discovered a real treasure trove of integrability with more than 50 new equations. 

The method used in this paper is a direct application of singularity confinement, whereupon we just decide which are the lengths of the singularity patterns. Of course, only the patterns that yield confined singularities are kept. Once the pattern is decided upon, writing the confinement conditions is elementary, as explained through the worked-out examples in section 4. The integration of these conditions leads to the precise parametrisation of the discrete Painlev\'e equation.

Since all equations obtained here are of  trihomographic form in each of the two dependent variables, it is possible to eliminate either of the two, obtaining a single equation. The most interesting result of this paper is that all equations obtained by such an elimination are among those derived in [\ancil]. This is true even for the equations which involve the rather unfrequently encountered periodic functions $\omega_6$ and $\omega_9$. This of course confirms that in deriving the results of [\ancil] no case was missed.  But what is perhaps more crucial is to ask whether all equations present in [\ancil] are present in the results of this paper. It turns out that this is true (taking into account that 5.2.9 is, in fact, identical to 5.2.10). What is also interesting is that for the patterns which lead to equations corresponding to an artificial asymmetrisation of cases already identified in [\second], namely (6,2,6,2), (4,4,4,4), (7,1,7,1) and (5,3,5,3), an elimination of either of the two variables leads to an equation corresponding to a double-step evolution as shown in [\eaone], [\restor] and [\sustem].

This paper was devoted to the study of additive equation exclusively. However this is not a limitation. Once a discrete Painlev\'e equation with E$_8^{(1)}$ symmetry is obtained in additive form, transcribing the results to the multiplicative and elliptic case is elementary, as we have shown in [\second]. That said, several extensions of the present work do suggest themselves. First, this study was limited to the asymmetric trihomographic case. One possibility would be to examine the non-trihomographic forms as was done in [\ancil] for symmetric systems. However one should keep in mind that going from trihomographic to non-trihomographic forms in that paper lead to an increase of an order of magnitude in the number of systems to be treated. Were the same scaling law to apply in this case, one would have to study more than 500 different equations, a task of herculean proportions. A different direction, and one that appears a priori more manageable, is to examine the possible limiting cases of our systems, just as was done in [\refdef\limits] for symmetric systems. We expect to be able to address this question is some future work of ours. 

\bigskip
{\scap References}
\medskip

[\dps] A. Ramani, B. Grammaticos and J. Hietarinta, Phys. Rev. Lett. 67 (1991) 1829.

[\jimbo] M. Jimbo and T. Miwa. Physica 2D (1981) 407.

[\fokas] A. Fokas, B. Grammaticos and A. Ramani, J. of Math. Anal. and Appl. 180 (1993) 342.

[\selfdu] A. Ramani, Y. Ohta, J. Satsuma and B. Grammaticos, Comm. Math. Phys. 192 (1998) 67.

[\sakai] H. Sakai, Commun. Math. Phys. 220 (2001) 165.

[\miura] A. Ramani and B. Grammaticos, Jour. Phys. A 25 (1992) L633.

[\fravas] F.W. Nijhoff and V. Papageorgiou, Phys. Lett. A 153 (1991) 337.

[\akns] M. J. Ablowitz, D. J. Kaup, A. C. Newell, and H. Segur,  Stud. Appl. Math.  53 (1974) 249.

[\disdres] B. Grammaticos, A. Ramani and V. Papageorgiou, Phys. Lett. A 235 (1997) 475.

[\shohat] J.A. Shohat, Duke Math. J. 5 (1939) 401.

[\brezov] E. Br\'ezin and V.A. Kazakov, Phys. Lett. 236B (1990) 144.

[\kanoya] K. Kajiwara, M. Noumi and Y. Yamada, J. Phys. A 50 (2017) 073001.

[\eight] Y. Ohta, A. Ramani and B. Grammaticos, J. Phys. A 34 (2001) 10523.

[\qrt] G.R.W. Quispel, J.A.G. Roberts and C.J. Thompson, Physica D34 (1989) 183.

[\sincon] B. Grammaticos, A. Ramani and V. Papageorgiou, Phys. Rev. Lett. 67 (1991) 1825.

[\procroy] T. Mase, R. Willox, B. Grammaticos and A. Ramani, Proc. Roy. Soc. A 471 (2015) 20140956.

[\full] A. Ramani, B. Grammaticos, R. Willox, T. Mase and M. Kanki, J. Phys. A 48 (2015) 11FT02.

[\trihom] B. Grammaticos and A. Ramani, J. Math. Phys. 56 (2015) 083507.

[\first] A. Ramani and B. Grammaticos, J. Phys. A 48 (2015) 355204.

[\second] B. Grammaticos and A. Ramani, J. Phys. A 48 (2015) 16FT02.

[\ancil] A. Ramani and B. Grammaticos, J. Phys. A 50 (2017) 055204.

[\multell] B. Grammaticos and A. Ramani, J. Phys. A 49 (2016) 45LT02.

[\eaone] B. Grammaticos, A. Ramani and R. Willox, {\sl Restoring discrete Painlev\'e equations from an E8-associated one}, preprint (2018) arXiv:1812.00712 [math-ph], to appear in J. Math. Phys..

[\restor] A. Ramani, B. Grammaticos, R. Willox and T. Tamizhmani, {\sl Constructing discrete Painlev\'e equations: from E8 to A1 and back}, preprint (2019) arXiv:1902.09920 [math-ph], to appear in J. Nonlin .Math. Phys.

[\sustem] R. Willox, A. Ramani and B. Grammaticos, J. Math. Phys. 58 (2017) 123504.

[\limits] K.M. Tamizhmani, T. Tamizhmani, A. Ramani and B. Grammaticos, J. Math. Phys. 58 (2017) 033506.

\end{document}